\documentclass[conference]{IEEEtran}
\IEEEoverridecommandlockouts
\usepackage{cite,url}
\usepackage{amsmath,amssymb,amsfonts}
\usepackage{algorithmic}
\usepackage{graphicx}
\usepackage{textcomp}
\usepackage{xcolor}
\def\BibTeX{{\rm B\kern-.05em{\sc i\kern-.025em b}\kern-.08em
    T\kern-.1667em\lower.7ex\hbox{E}\kern-.125emX}}
\begin{document}

\title{A context-aware multiple Blockchain architecture for managing low memory devices\\
}

\author{\IEEEauthorblockN{Marco Fiore}
\IEEEauthorblockA{\textit{Department of Electrical and Information Engineering} \\
\textit{Polytechnic University of Bari}\\
Bari, Italy \\
marco.fiore@poliba.it}
\and
\IEEEauthorblockN{Marina Mongiello}
\IEEEauthorblockA{\textit{Department of Electrical and Information Engineering} \\
\textit{Polytechnic University of Bari}\\
Bari, Italy \\
marina.mongiello@poliba.it}
\and
\IEEEauthorblockN{Giuseppe Acciani}
\IEEEauthorblockA{\textit{Department of Electrical and Information Engineering} \\
\textit{Polytechnic University of Bari}\\
Bari, Italy \\
giuseppe.acciani@poliba.it}
}

\maketitle

\begin{abstract}
Blockchain technology constitutes a paradigm shift in the way we conceive distributed architectures. A Blockchain system lets us build platforms where data are immutable and tamper-proof, with some constraints on the throughput and the amount of memory required to store the ledger. This paper aims to solve the issue of memory and performance requirements developing a multiple Blockchain architecture that mixes the benefits deriving from a public and a private Blockchain. This kind of approach enables small sensors - with memory and performance constraints - to join the network without worrying about the amount of data to store. The development is proposed following a context-aware approach, to make the architecture scalable and easy to use in different scenarios.
\end{abstract}


\section{Introduction}
% Introduzione
Blockchain technology - a form of Distributed Ledger Technology (DLT) - can be used to securely store information. It can be represented as a distributed database consisting of a chain of logically linked blocks, each containing multiple transactions. Blockchain is an add-only data storage solution: data can only be added and concatenated to previous one, so tampering becomes very difficult since modifying a single block can involve breaking the entire chain. Use of Blockchain is recommended when the integrity of the archived data is threatened by the existence of multiple users capable of simultaneously accessing data. The most popular applications of the Blockchain platform regard cryptocurrencies and food traceability.

% Related
There are several approaches that can be analyzed towards Blockchain-based traceability architectures, ranging from agri-food \cite{shew2022consumer, dasaklis2022systematic, feng2020applying, tian2016agri, caro2018blockchain} to healthcare \cite{omar2022blockchain, kamath2022blockchain, yaqoob2022blockchain} and smart cities \cite{fiore2023iccsm}. Most of the analyzed architectures propose an approach that takes advantage of Internet of Things (IoT) devices \cite{balamurugan2022iot, ali2017iot}: in these cases Blockchain and Distributed Ledger Technology ensure availability and traceability of data exchanged and stored in the Supply Chain. Other implementations consider just one kind of Blockchain architecture, either public \cite{cocco2021blockchain} or private \cite{uddin2021blockchain, gao2020design}, even in conjunction with the Inter Planetary File System (IPFS) \cite{huang2019food}.\\

A common issue in these approaches is to guarantee high performance even though the usage of devices having a small amount of memory and poor performance. A Blockchain system requires devices with good memory, but no implementations are proposed on the application of a private Blockchain together with a public one to correctly manage data coming from sensors. Consistency of data is solved using a common interface for the two ledgers: information is exchanged in JSON format, both as output from the private side and as input for the public side.\\

% Scopo
The main goal of this paper is the proposal of a new architecture in which multiple Blockchains are combined to guarantee good performances using low storage and performances devices like Internet of Things (IoT) sensors. To this purpose, a context-aware smart contract application has been developed. Such approach guarantees the possibility to ensure that even data gathered from IoT sensors are immutable, even if some considerations on the limitations of Blockchain technology in this topic (size, throughput, real-time data analysis) must be done. A private Blockchain is used as a secure and immutable middle layer between the IoT sensor, responsible of uploading data, and the public Blockchain, responsible of gathering value extracted from data and let this new information spread in the network. To sum up, we describe how the system works through a scenario and outline strengths and limitations of the approach.
\section{Blockchain technology}
\begin{figure}[!ht]
	\centering
	\includegraphics[width=.9\linewidth]{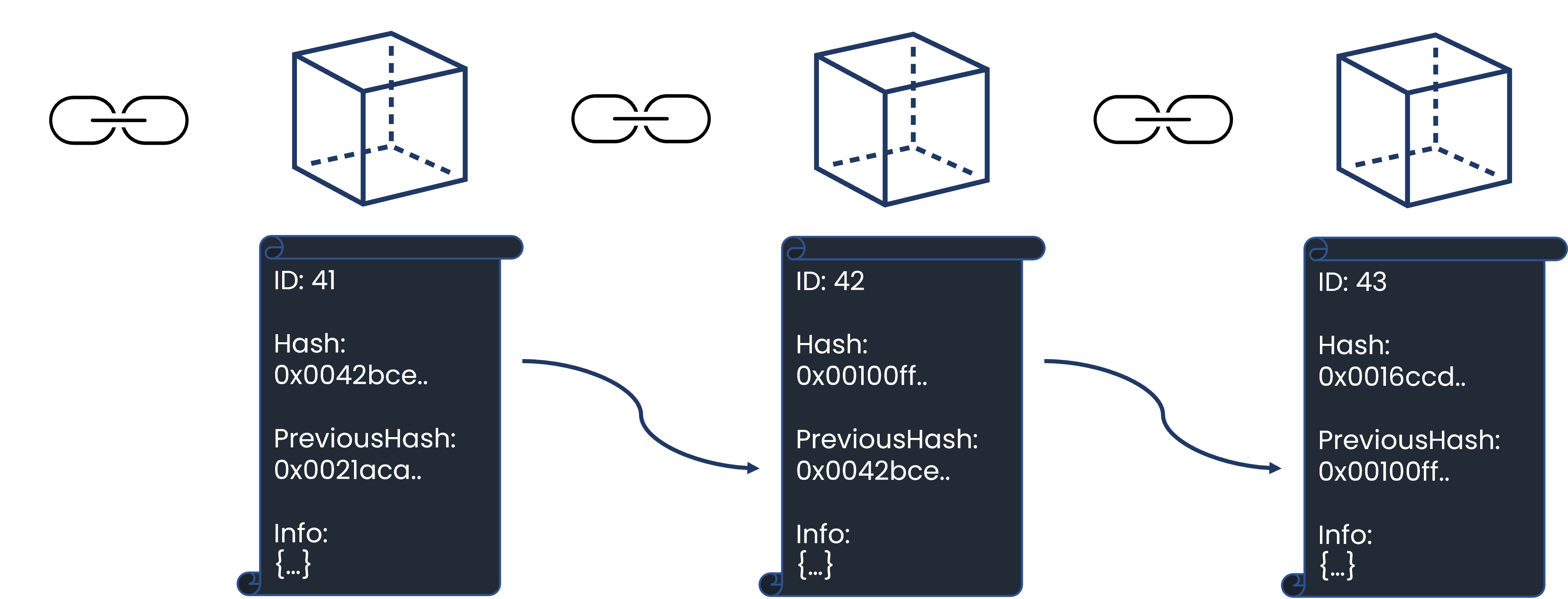}
	\caption{Simple Blockchain architecture: blocks are connected one to the following through the "PreviousHash" field.}
	\label{fig:blockchain}
\end{figure}

Blockchain literally means chain of blocks: its main peculiarity is the decentralization and distribution of a database on multiple devices. A sample Blockchain is shown in Fig. \ref{fig:blockchain}. Blockchain and Distributed Ledgers are not based on centralized systems: this means that data are not stored in a single place (i.e., a server) and controlled and shared by a single entity. In a distributed ledger, each node in the network holds a copy of the ledger: changes in data are broadcasted to every participant. In this way, Blockchain becomes trustable and not dependable on central authorities.\\

In a Blockchain platform, each transaction between two participants is permanently recorded. These records are named blocks, whereas each computer used for processing Blockchain is called node. A transaction can be added through a mining operation, a process that adds one or more records to the Blockchain. The mining process is based on hash functions that are computationally complex. Encryption of the shared information carries a high level of security. Changes must be accepted with the public consensus.\\

There are three types of Blockchain, named public Blockchain, private Blockchain and consortium or federated Blockchain. Their main features are summarized in Table \ref{tab:blockchain_types} \cite{zheng2017overview}.
\begin{itemize}
    \item Public Blockchain\\
    A public Blockchain has not nodes that control the network; everyone can join the distributed ledger and insert or read the information stored in it. A public Blockchain is open and transparent; algorithms used to support consensus are Proof-of-Work, Proof-of-Stake or Proof-of-Authority.
    \item Private Blockchain\\
    A private Blockchain is managed by an organization or an individual who gains control over the network. Mining rights can be given to anyone, but the decision is taken by the organization. In this case, the ledger can be considered as more centralized with respect to the public Blockchain since there is a single entity that owns more rights than others. A private Blockchain is less expensive than a public one.
    \item Consortium Blockchain\\
    A consortium Blockchain tries to mix benefits of public and private Blockchains. In this case, the owners of the ledger are multiple nodes instead of a single one. This makes the Blockchain more decentralized than the private one, but not as expensive as the public one. The group of owners manages the Blockchain and keeps the network alive.
\end{itemize}

\begin{table}[htbp]
\caption{Comparison between public, private and consortium Blockchain.}
\centering
\resizebox{.45\textwidth}{!}{%
\begin{tabular}{|l|l|l|l|}
\hline
& \multicolumn{1}{c|}{\textbf{Public}} & \multicolumn{1}{c|}{\textbf{Private}} & \multicolumn{1}{c|}{\textbf{Consortium}} \\ \hline
\textbf{Consensus determination} & All miners                           & One organization                      & Selected set of nodes                    \\
\textbf{Read permission}         & Public                               & Public or restricted                  & Public or restricted                     \\
\textbf{Immutability}            & Impossible to tamper                 & Could be tampered                     & Could be tampered                        \\
\textbf{Efficiency}              & Low                                  & High                                  & High                                     \\
\textbf{Centralised}             & No                                   & Yes                                   & Partial                                  \\
\textbf{Consensus process}       & Permissionless                       & Permissioned                          & Permissioned                             \\ \hline
\end{tabular}%
}
\label{tab:blockchain_types}
\end{table}

\section{Proposed architecture}
\begin{figure*}[htbp]
    \centering
	\includegraphics[width=\textwidth]{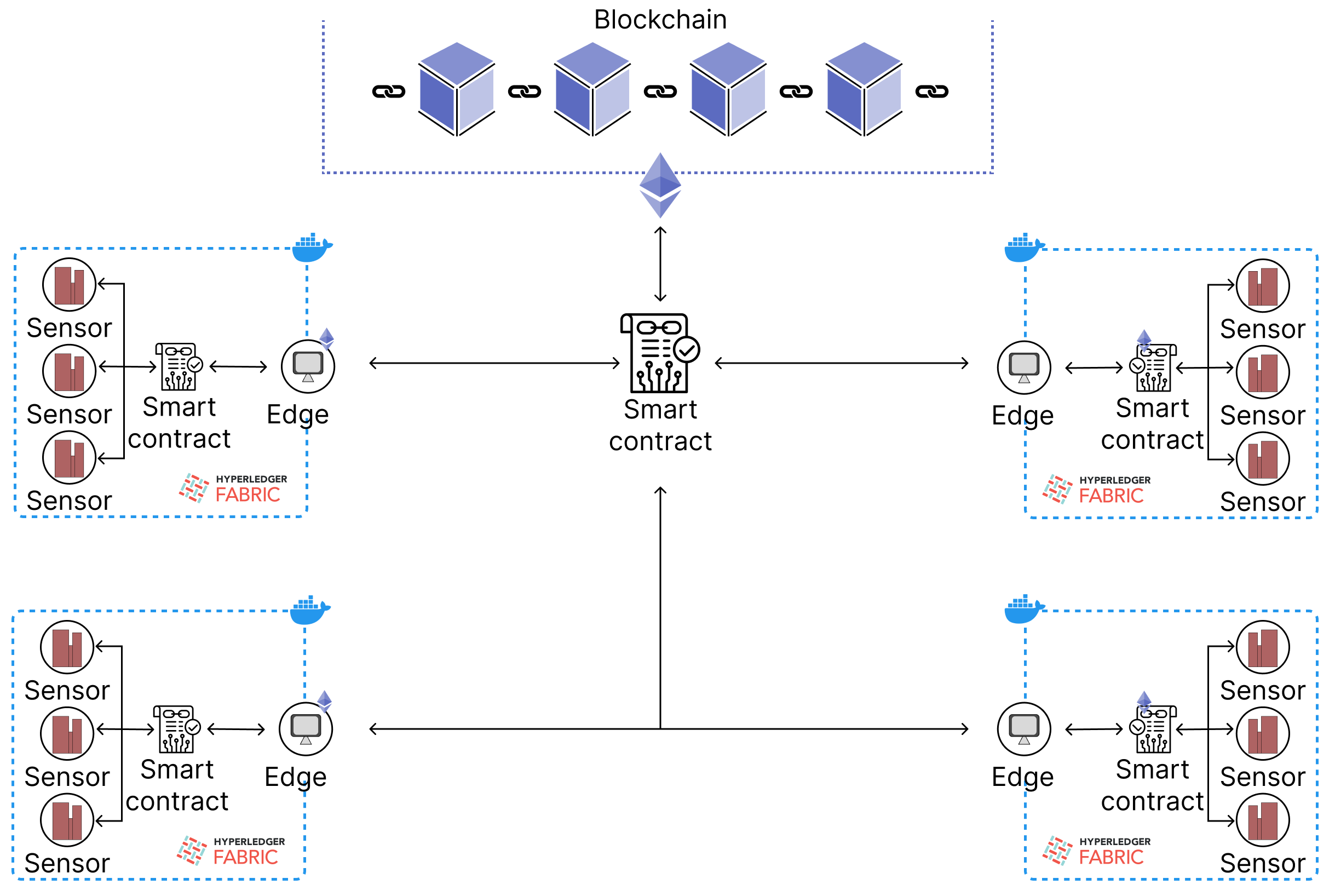}
	\caption{Architecture of the proposed platform}
	\label{fig:proposed_architecture}
\end{figure*}

Our proposed architecture is shown in Fig. \ref{fig:proposed_architecture}. It is mainly made up of (a) containers (for the implementation of Hyperledger Fabric), (b) smart contracts, (c) edge computing units and (d) IoT sensors. Here, the interaction between different Blockchains is highlighted. The main system is based on Ethereum\footnote{\url{https://ethereum.org/en/}}, to guarantee all the benefits of using a public Blockchain: transparency, data immutability, reliability. The Ethereum nodes consist of some edge computing units, responsible for management of all the internal low-memory devices. These devices will not rely on some NoSQL databases as described by authors of papers \cite{ge2022hybrid, nasreen2022implementation} to communicate and store data: they use a containerized private Blockchain, in particular we use Hyperledger Fabric\footnote{\url{https://hyperledger-fabric.readthedocs.io/en}} but to this purpose any private Blockchain can be used. This approach provides modularity and scalability to the platform. Each edge computing unit is at the same time an Hyperledger peer and an Ethereum node: from Hyperledger side, it can obtain data from IoT sensors, manage and convert them into valuable data; from Ethereum side, it can upload the output of the computation (i.e. the mean and standard deviation of monthly data) lightening the Blockchain load. Due to their limited memory and computational power, IoT sensors can only benefit from the application of our methodology: sensor's ledger is less expensive than the Ethereum one, because they only keep track of the information collected inside their container. The behaviour of a container is independent of all others, thus ensuring modularity and scalability to the platform.

\begin{figure}[!ht]
\centerline{\includegraphics[width=.8\linewidth]{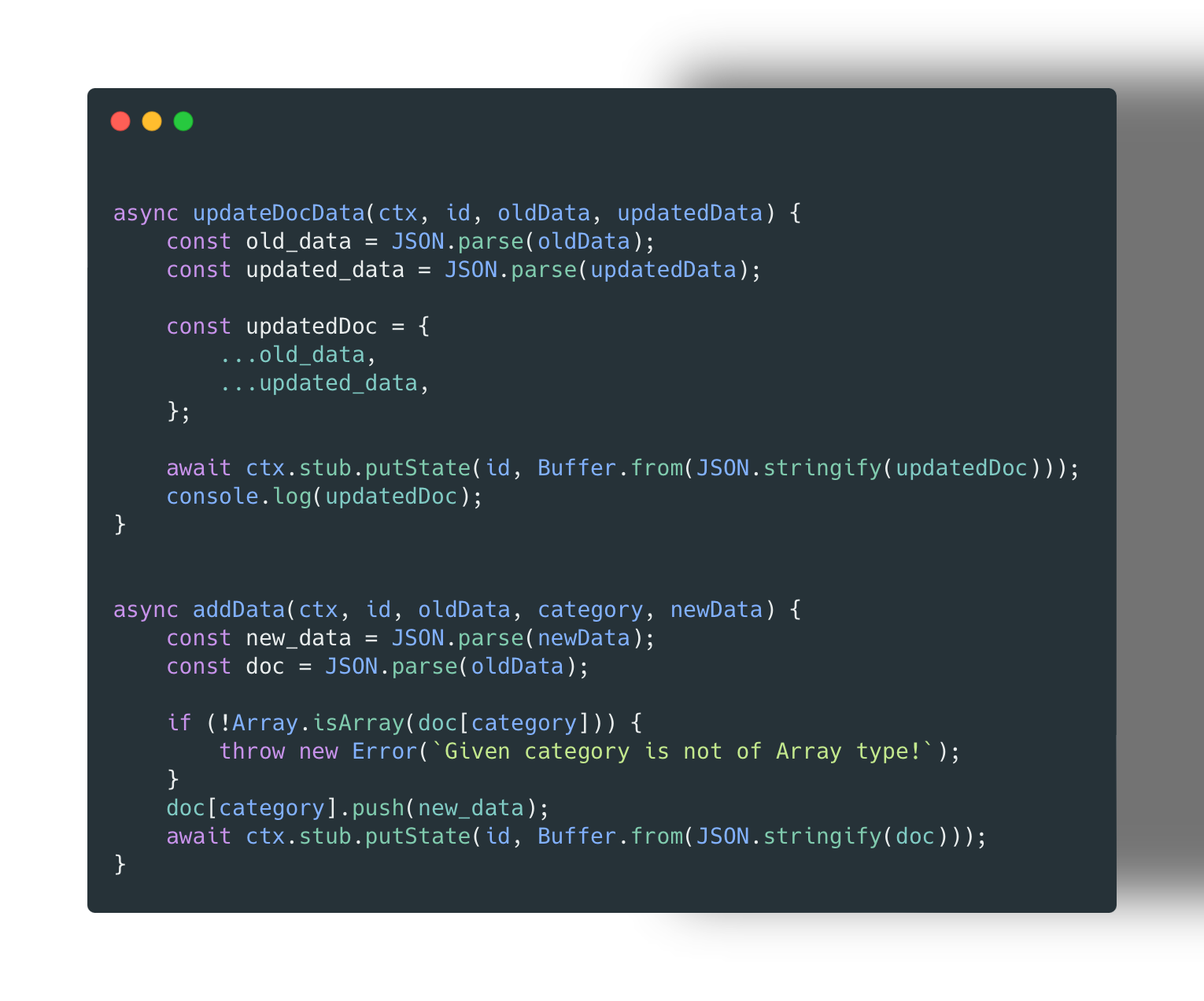}}
\caption{Hyperledger Fabric chaincode written in a context-aware approach.}
\label{fig:api}
\end{figure}

\begin{figure}[htbp]
\centerline{\includegraphics[width=.9\linewidth]{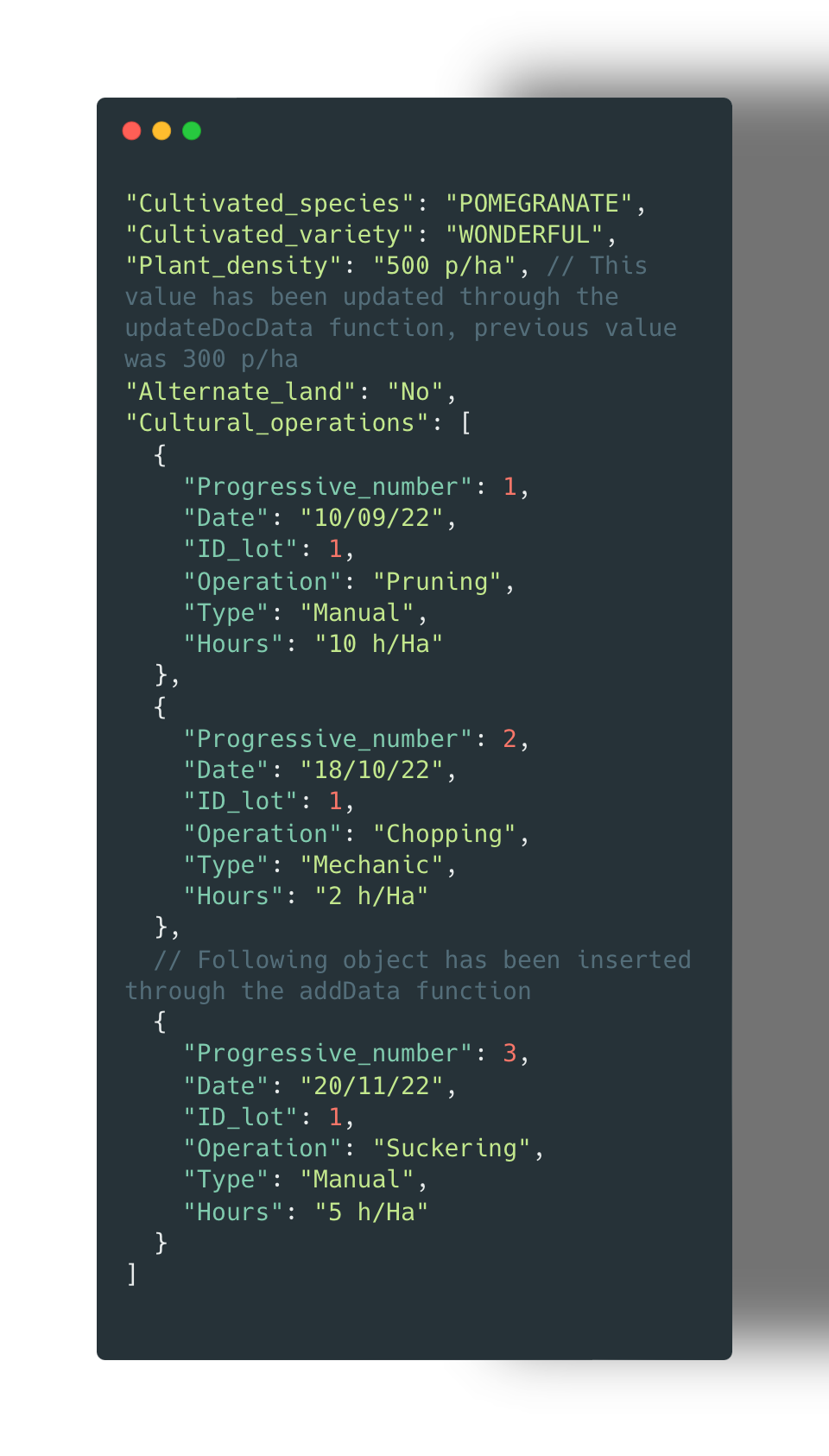}}
\caption{Execution of the Hyperledger Fabric chaincode in a Blockchain-based traceability system.}
\label{fig:hf_execution}
\end{figure}

Smart contracts are used to manage the public and private Blockchain. On Hyperledger side, they are developed following a context-aware approach: they can receive any kind of input data from sensors and insert them into the database using a JSON format. Fig. \ref{fig:api} shows the JavaScript chaincode for updating data or adding new information to an array. Fig. \ref{fig:hf_execution} shows the output of such functions in a food traceability system with an updated value (Plant density) and a new information inserted in the Cultural Operations array. The edge computing unit reads the inserted data, converts them in valuable data and sends the final information to Ethereum. In this way, thanks to the applied algorithm, the amount of data sent to Ethereum and stored in Hyperledger Fabric subsystem is reduced. Consistency of data is not a problem in this approach, because a common interface is used: information between platforms is exchanged using a JSON format, that is managed by the JavaScript chaincode and by the Solidity smart contract.

\section{Scenario}
The proposed system has been implemented following a scenario: design and development of a traceability system for agri-food supply chain. Our schema is different with respect to the approaches described in papers \cite{madine2021appxchain, jiang2019cross, oktian2020hierarchical}.

The scenario proposes five different fields, each of them covering an agri-food product: asparagus, pomegranate, almond, tomato and durum wheat. The system is split into three main parts: collecting data, transforming data into value, sending value to the public Blockchain.

\subsection{Collecting data}
Data collection is possible thanks to some IoT sensors in the field that measure temperature and humidity and meteorologic parameters, ranging from the rain percentage during the day to the wind speed. These data are not valuable if considered standalone, but they should be gathered to extract useful information. Furthermore, these big data occupy lots of memory, so it would be impossible for a low memory device to obtain an exhaustive storage of the information coming from all the sensors. By creating a local private Blockchain, devices can easily manage smaller amounts of data. The usage of a private Blockchain, with respect to a general database such as MongoDB, grants immutability of data even in this phase of the process. No one can alter data, so counterfeiting is avoided.

\subsection{Transforming data into value}
Each field is gifted with an edge computing unit, that is responsible of reading every information that the sensors store in the private Blockchain. These units take data and convert them into value, performing some preliminary analysis on the importance and priority of the gathered information. The output of this process is a lightweight set of information that can be stored in the public Blockchain, to make them accessible by anyone.

\subsection{Sending value to the public Blockchain}
Every edge computing unit is a peer in the private Blockchain and a node in the public one. Only these units can upload new information to the public Blockchain, so only the final reports of each field is accessible to people. This process makes the traceability system transparent and generates trust in the final user. The consumer, before buying an agricultural product in the supermarket, can read information about a product by scanning a QR code to know where the product has been produced, the temperature of the field, the number of cultural operations.

\subsection{Strengths and limitations}
The proposed architecture is simple to implement in new and existing systems, because it requires few hardware elements to work correctly. A single edge computing unit is enough to create the platform, because it can be connected to both Hyperledger Fabric (for IoT devices) and Ethereum (for data uploading). Modularity is intrinsic in the platform: each set of IoT devices can create a new Hyperledger Fabric image, using container-based approaches such as Docker. This method allows high performances and scalability in scenarios involving many IoT devices. Also security is important: in this architecture IoT devices do not contain any sensitive information, even though they are frequently and easily subject to attacks. In fact, data processing is performed by the edge computing unit.\\
In addition to this, thanks to smart contracts, it is possible to provide controls that automatically exclude from the analysis and the processing phase out-of-scale values due to malfunction or tampering with the IoT sensors.

Sensitive data are generated by the edge computing unit and inserted directly into Ethereum: this procedure ensures transmission security due to the Blockchain, but at the same time makes this unit become a point of failure of the architecture. Anyway, the addition of an edge computing unit allows data redundancy and avoids malfunctions especially if located in a different place. From the security point of view, it is possible to reduce the number of accesses to the system to a minimum and to avoid human suspicious accesses.

\begin{figure*}[htbp]
\centerline{\includegraphics[width=\linewidth]{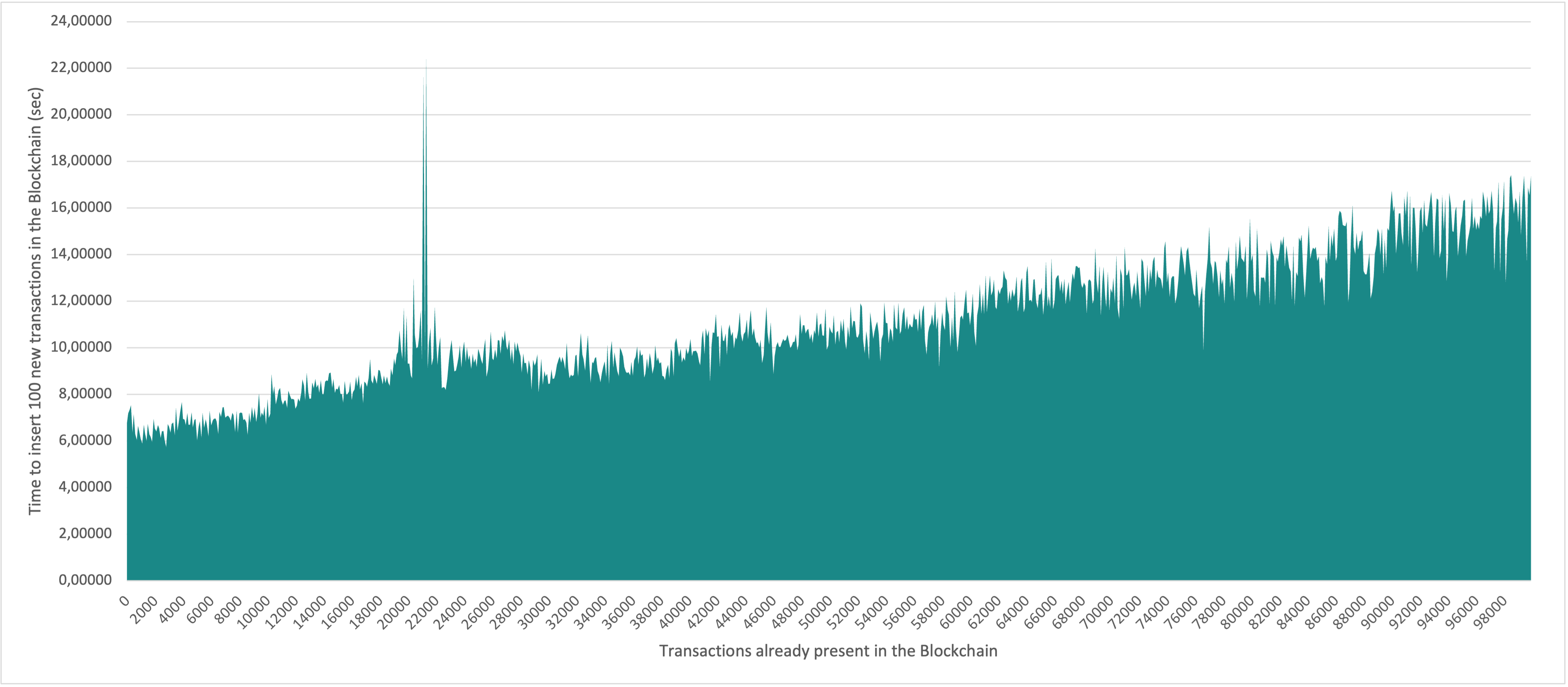}}
\caption{Time to manage 100 new values when N values are already present in the Blockchain.}
\label{fig:time}
\end{figure*}

Blockchain, as explained before, is an add-only storage solution, but we are using it in a scenario composed by low-memory devices. Fig. \ref{fig:time} shows the weight of a private Blockchain when N values are already written and a batch of 100 new values is inserted, while Table \ref{tab:validation-memory} shows the memory occupied by transactions in the private Blockchain and the time needed to add a new batch of 100 transactions. To pursue our goal, the private Blockchain (Hyperledger Fabric) can gather data in a limited amount of time (i.e., a month), then perform a dimension reduction to collected data, upload it in Ethereum and discard old information (i.e., resetting the private ledger). This is not an issue for our proposal because data is still present in the system: its value has been already uploaded in the public Blockchain, so single data are not needed anymore.

\begin{table*}[htbp]
\caption{Memory needed to store transaction in the Blockchain and time needed to upload 100 new transactions}
\centering
\begin{tabular}{|c|c|c|}
\hline
\textbf{Transactions in Blockchain} & \textbf{Occupied storage (MB)} & \textbf{Time needed to add 100 transactions {[}sec{]}} \\ \hline
0      & 0,219 & 0        \\ \hline
5      & 0,272 & 0,36     \\ \hline
10     & 0,339 & 0,84     \\ \hline
50     & 0,635 & 4,44     \\ \hline
100    & 0,892 & 8,46     \\ \hline
500    & 2,5   & 40,26    \\ \hline
1000   & 6,3   & 86,46    \\ \hline
5000   & 32,4  & 421,26   \\ \hline
10000  & 64,5  & 912,66   \\ \hline
50000  & 335,7 & 5508,66  \\ \hline
100000 & 670,6 & 14688,66 \\ \hline
500000 & 3400  & 62688,66 \\ \hline
\end{tabular}
\label{tab:validation-memory}
\end{table*}

\section{Conclusion}
In this paper, we have proposed a multiple Blockchain approach that is feasible to multiple scenarios thanks to context-aware smart contracts. The framework is based on a modular architecture capable of supporting low memory devices without lacking on security. The proposed scenario is based on agri-food traceability and shows that this kind of approach is useful when it comes to modularity: it is possible to easily add new fields and new IoT devices with containerized private Blockchains such as Hyperledger Fabric, using a single edge computing unit to gather and convert data and to upload new information to Ethereum. The implementation of two different Blockchains lets the user build a secure infrastructure, where data is immutable since its creation through an IoT device to its conversion through the edge computing unit to its final upload in the public ledger. To further improve security, some additional edge computing units can be considered for backup purposes, or it is possible to delegate data conversion to specific smart contracts as described in paper \cite{kushwaha2022systematic}, thus finding the trade-off between performances and reliability.

\bibliographystyle{ieeetr}
\bibliography{refs}

\end{document}